\documentclass[12pt,twoside,a4paper,fleqn]{article}
\usepackage[left=25mm,top=20mm,right=14mm,bottom=25mm]{geometry}

\usepackage{epsfig}
\usepackage{graphicx}
\usepackage{amssymb}
\usepackage{mathrsfs}
\usepackage{dcolumn}

\usepackage{multirow}

\newcommand{\pr}{\partial}

\newcommand{\rta}{\rightarrow}

\newcommand{\upa}{\uparrow}
\newcommand{\dwa}{\downarrow}
\newcommand{\ep}{\epsilon}
\newcommand{\ve}{\varepsilon}
\newcommand{\p}{\prime}
\newcommand{\om}{\omega}
\newcommand{\ra}{\rangle}
\newcommand{\la}{\langle}

\newcommand{\intin}{\int_{-\infty}^{+\infty}}
\newcommand{\beq}{\begin{equation}}
\newcommand{\eeq}{\end{equation}}

\newcommand{\ball}{\begin{align}}
\newcommand{\eall}{\end{align}}

\newcommand{\beqar}{\begin{eqnarray}}
\newcommand{\eeqar}{\end{eqnarray}}

\newcommand{\dg}{\dagger}

\newcommand{\ben}{\begin{enumerate}}
\newcommand{\een}{\end{enumerate}}
\makeatletter
\newcommand*{\rom}[1]{\expandafter\@slowromancap\romannumeral #1@}
\makeatother

\begin{document}
\title{Resistivity of a $2d$ quantum critical metals}
\author{ Komal Kumari$^1$, Raman Sharma$^1$ and Navinder Singh$^2$ \\ $^1$Department~of~Physics,~Himachal~Pradesh~University,\\~Shimla,~India, Pin:171005.\\$^2$Physical Research Laboratory, Ahmedabad,\\ India, Pin: 380009.
\footnote{~Email:~sharmakomal611@gmail.com,}
\footnote{~Email:~raman.sharma@hpuniv.ac.in, }
	\footnote{ ~Email:~navinder@prl.res.in}}
\maketitle
\begin{abstract}
We calculate resistivity in the paramagnetic phase just above the curie temperature in a $2d$ ferromagnetic metal. The required dynamical susceptibility in the formalism of resistivity is calculated within the Random Phase Approximation(RPA). The mechanism of resistivity is magnetic scattering, in which $s$-band electrons are scattered off the magnetic spin fluctuations of d-band electrons. We use the $s$-$d$ Hamiltonian formalism. We find that near the quantum critical point the resistivity in $2d$ scales as $T^{\frac{4}{3}}$, whereas in $3d$ it scales as $T^{\frac{5}{3}}$. In contrast to it, resistivity due to phonon scattering is given by  $T^5$ in low temperature limit as is well known. Our RPA result agrees with the Self-Consistence Renormalisation(SCR) theory result.
\end{abstract}

\section{Introduction}
Resistivity in metals is generally due to either impurity scattering or phonon scattering or both mechanism working together. In impurity scattering electrons are scattered off the immobile impurities leading to their momentum randomization thus resistivity. In phonon scattering electrons scatter by absorbing or emitting phonons or lattice vibrations\cite{singh}. This can be studied using Bloch-Boltzmann kinetic equation and one finds that resistivity is proportional to temperature($T$)  when $T\gg\Theta_{D}$, where $\Theta_{D}$ is the Debye temperature. In the opposite limit $T\ll\Theta_{D}$, $\rho\propto T^5$.\\

However, the above scenario is not applicable to magnetic materials tuned near their critical points. An alternative mechanism in which electrons scatter off magnetic spin fluctuations\cite{mathon} becomes important. Currently, there is a renewed interest in the topic of magnetic critical points and physical properties near a magnetic instability\cite{belitz,stewart,sachdev,olivier, varma,gonzalo,senthil,abanov,liam,takashima,takahashi,moriya}.  It has been shown that electron-magnetic-spin-fluctuation scattering in a $3d$ ferromagnetic metal\cite{mathon} tuned near to its critical point leads to a resistivity which scales as $T^\frac{5}{3}$. This stands in sharp contrast to phonon scattering. It has also been shown within the Self-Consistence Renormalisation(SCR)\cite{takahashi,moriya,kawabata,takimoto} theory that in $2d$ ferromagnetic case resistivity scales as $T^\frac{4}{3}$ near the critical point. In this paper we represent our calculation of resistivity in a $2d$ ferromagnetic metal near its critical point by using Random Phase Approximation(RPA) instead to SCR theory. We report that our RPA result agree with the SCR theory result, that is  $\rho \propto T^\frac{4}{3}$.\\

One importance of our result is that the finite temperature renormalization effects taken into account in SCR theory are not important near a quantum critical point when one study transport properties. And RPA is as good as SCR theory(at least in the case of electrical resistivity) in this very low temperature regime near $QCP\cite{kawabata,takimoto}$.\\

Our calculation (within the $s-d$ Hamiltonian formalism) is applicable to weakly ferromagnetic alloys such as Ni-Pd which has certain amount of quenched disorder. This is important as quenched disorder can lead to second-order phase transition near the critical point\cite{kirkpatrick}. Our calculation is for second order ferromagnetic phase transition. In clean ferromagnets the presence of fermionic soft modes makes the transition first order as there is an extra entropy associated with a collective excitations just above the ferromagnetic transition\footnote{If enegry of a fermionic mode is $\hbar \om_{q}$ and density is $n_{q}$, then the extra energy needed to excite soft modes can be roughly given by $\int_{0}^{\infty}d^3 q ~n_{q}\hbar\om$ and entropy is $\frac{1}{T}\int_{0}^{\infty}d^3 q ~n_{q}\hbar\om $. }. It turns out that in the presence of disorder these fermionic soft modes becomes diffusive and the transition can become second order. For more details refer to \cite{kirkpatrick}.\\
In the next section we present the formalism and our calculation of resistivity.

\section{Formalism}
 $s$-electrons are treated as conduction electrons which scatter via localised $d$-electrons, and their interaction is modelled with $s$-$d$ Hamiltonian\cite{mathon}:

\beqar
H_{int}=\frac{J}{N}\sum_{k,k^\p} \bigg\{ a^{\dagger}_{k^{\p}\uparrow} a_{k\downarrow}S^{-}(k^{\p}-k)+ a^{\dagger}_{k^{\p}\downarrow} a_{k\uparrow}S^{+}(k^{\p}-k)+(a^{\dagger}_{k^{\p}\uparrow}
 a_{k\upa}-a^{\dagger}_{k^{\p}\downarrow} a_{k\dwa})S^{z}(k^{\p}-k)\bigg\} \label{h1}
\eeqar
where $J$ is $s$-$d$ electron coupling constant, $N$ is the number of atoms in the system, the $a^\dg$ and $a$ are the creation and annihilation operators for $s$-electrons. $S^{-}(k^{\p}-k)$ and $S^{+}(k^{\p}-k)$ are lowering and raising spin density operators of $d$-band electrons, $S^{z}(k^{\p}-k)$ represents $z$-component of spin density of $d$-band electrons and  $S(k)$ is the Fourier transform of spin density $S(r)$ of d-electrons and it is defined by 
\beq
S(k)=\int e^{-ik.r} S(r)dr \label{h2}
\eeq
 
The transition probability\cite{mills}  that an  $s$-electron with wave vector $k$ will be scattered to be the state $k+q$ can be written as\\
\beqar
W_{k+q \leftarrow k}= \frac{2\pi}{\hbar}|\la d(F)|\la k+q|H_{int}|k\ra| d(I)\ra|^2 \rho_{f}. \label{h3}
\eeqar
The eigenstate of the system can be approximated by the product of the form $|k\ra| d(I)\ra$, where the function $|k\ra$ describes the state of the $s$-electrons, and $| d(I)\ra$ the states of the $d$-electrons system. By employing the Fermi Golden Rule, one finds that $s$-electron of wave vector $s$ with spin up$/$down is scattered into the state $k+q$ with spin down$/$up at the rate\cite{mills}
\beqar
W^{para}_{k+q \leftarrow k}&=&\frac{2\pi J^2}{\hbar N^2}f_{s}(\ep_{k})(1-f_{s} (k+q))\sum_{I,F}\bigg\{<d(I)S^{+}(-q)|d(F)><dF|S^{-}(q)|d(I)>+\nonumber\\&&~<d(I)|S^{-}(q)|d(F)><d(F)|S^{+}(q)|d(I)>\bigg\}\delta\bigg(\ep_{d}(F)-\ep_{d}(I)+\ep_{s}(k+q)-\ep_{s}(k)\bigg).\nonumber\\&& \label{h4}
\eeqar

where $f_{s}(\ep_{k})$, is $s$-electron Fermi distribution function. The last term in Hamiltonian from the $z$-component of spin density of $d$-electrons get cancelled due to scattering from the  same spin state of $s$-electrons. We employ the identity $\delta{\ep}=\int e^{i\ep t}\frac{d t }{2\pi}$ and  $e^{-i H_{d}(I)} |d(I)>=e^{-i \ep_{d}(I)}|d(I)>$. Therefore we obtain:

\beqar
W^{para}_{k+q\leftarrow k}&=& \frac{3}{2}\frac{J^2}{\hbar N^2} f_{s}(\ep_{k})(1-f_{s}(\ep_{k+q})) \int_{0}^{\infty} dt e^{-i\om t}[\la S^{+}(-q)S^{-}(q,t)\ra +\la S^{-}(-q)S^{+}(q,t)\ra] \nonumber\\ \label{h5}
\eeqar

where $\hbar\om=\ep_{k}-\ep_{k+q}$ is an enegry transfer provided by $s$-electrons, $\ep_{k}=\frac{\hbar^2 k^2}{2m_{s}}$, and $<...> $ denotes the thermal average. Time evolution of operators implies Heisenberg representation: $S^{\pm}(q,t)=e^{iH_{d}t}S^{\pm}e^{- i H_{d}t}$. The intergral term describes the Fourier transform of the correlation function of spin densities of $d$-band electrons.  Using the Fluctuation-dissipation theorem\cite{izuyama}, we express the transition probability $W^{para}_{k+q\leftarrow k}$ in terms of the dynamical susceptibilty $\chi^{-+}(q,\om) $ as

\beqar
W^{para}_{k+q \leftarrow k}= \frac{3}{2} 4\frac{J^2}{ N^2}\intin d\om f_{s}(\ep_{k})(1-f_{s}(\ep_{k+q}))(-n(-\om)) Im\chi_{s}^{-+}(q,\om) \delta(\hbar\om-\ep_{k}+\ep_{k+q})\label{h6}
\eeqar
where $n(-\om)=\frac{1}{e^{-\beta\hbar\om}-1}$ and $\chi^{-+}_{s}(q,\om)$ is the symmetric part of the complex susceptibility corresponding to a magnetic scattering of spins of d-band electrons with wave vector $q$ and frequency $\om$. $\chi^{+-}({q,\om})$ the susceptibility of d-electrons is defined by\\
\beqar
\chi^{+-}(q,\om)=\lim_{\ve \to 0}\frac{i}{\hbar}\int_{0}^{\infty} e^{-i\om t-\ve t}\la[ S^{+}(q,t),S^{-}(-q)]\ra d t \label{h7}
\eeqar
This dynamical susceptibility is the Fourier transform of response function or retarded Green function defined with respect to spin densities of d-electrons post scattering. The susceptibility tensor is  isotropic for paramagnetic system, therefore omitting anisotropy for the present system
\beq
\chi^{+-}(q,\om)=\frac{2}{3}(\chi^{xx}+\chi^{yy}+\chi^{zz}) \label{h8}
\eeq

The susceptibility explicitly defines that in an isotropic paramagnetic system $s$-electrons are equally scattered from all three components of magnetization.
 \par
The standard transport theory\cite{ziman} can now be used to calculate the transport property in terms of scattering probability $W_{k+q \leftarrow k}$
\beqar
\rho^{para}= \frac{1}{2 k_{B} T}\frac{\int\int(\Phi_{k}-\Phi_{k+q}) W^{para}_{k+q\leftarrow k}}{|\int e v_{k}\Phi_{k}\frac{\partial f^{0}(\ep_{k})}{\partial \ep_{k}} d k |^2 } d \mathbf{k} d \mathbf{q}  \label{h9}
\eeqar
where e is the electronic charge, $v_{k}=\frac{\hbar k}{ m_{s}}$ is the Fermi velocity of s-electrons, and  $-\Phi_{k}\frac{\partial f^{0}(\ep_{k})}{\partial \ep_{k}}$ is the measure of deviation from the equilibrium in the electron distribution, the trial function $\Phi_{k}$ itself a measure of this deviation. If the usual assumption $\Phi_{k}=const. \times q.u$ is made, and the variational integral in the denominator is solved by the assumption of isotropy in the electron distribution, then the resistivity expression  (\ref{h9}) reduces to 

\beqar
\rho^{para}_{2d}&=&\frac{3		
	J^2 \hbar^2}{ N^2 k_{B} T (en_{s})^2} \intin d\om \int_{0}^{K_{F}} d^2 \mathbf{k} \int d ^2\mathbf{q}  (u.q)^2  f^{0}(\epsilon_{k})(1-f^{0}(\ep_{k+q})) (-n(-\om))\nonumber\\&&
~~~~~~~~~~~~~~~~~~~~~~~~~~Im \chi^{-+}(q,\om) ~\delta(\hbar \om-\ep_{k}+\ep_{k+q})  \label{h10}
\eeqar

where u is a unit vector parallel to the eletric field and n is the number of s-electrons per unit volume. Using property $f(x)\delta(x-a)=f(a)\delta(x-a)$, and writing $\int d^2\mathbf{k}=\int_{0}^{\infty} kdk\int_{0}^{2\pi} d\phi$
\beqar
\rho^{para}_{2d}&=&\frac{3( \pi)  J^2 \hbar^2}{ N^2 k_{B} T (e n_{s})^2} \intin d\om ~ (-n(-\om))  \int q^3~ d q Im \chi^{-+}(q,\om) \int_{0}^{\infty} k ~ d k \times
\nonumber\\&& ~~~~~~~~~~~~~~~~~~f^{0}(\epsilon_{k})(1-f^{0}(\ep_{k}-\hbar \om))    \int_{0}^{2\pi}d\phi\delta(\hbar \om-\ep_{k}+\ep_{k+q})\nonumber\\ \label{h11}
\eeqar
Simplifying the integral with respect to $\phi$(appendix A) and writing $(u.q)^2=\frac{q^2}{2}$ for unit vector $u$, which is parallel to electric field and $q$. The expression gives
\beqar
\rho^{para}_{2d}&=&\frac{3(2 \pi)  J^2 }{ N^2\hbar^2 k_{B} T (e n_{s})^2} \intin d\om ~(- n(-\om))  \int q^2~ d q Im \chi^{-+}(q,\om)\times~\nonumber\\ &&~~~~~~~~~~~~~~~~~~~~~~~~~~~~~~~~~~ \int_{q_{0}}^{\infty} \frac{k ~ d k 
f^{0}(\epsilon_{k})(1-f^{0}(\ep_{k}-\hbar \om))}{\sqrt{k^2-q_{0}^2(q,\om)}}    \nonumber\\ \label{h12}
\eeqar

here $q_{0}=\frac{q}{2}+\frac{m \om}{\hbar q}<k_{F}$, $k_{F}$ is Fermi wave vector of $s$-electron and on performing k integral(appendix B) expression $(\ref{h12})$ reduces to
\beqar
\rho^{para}_{2d}&=&\sqrt{\frac{m}{2\mu}}\frac{3(2 \pi)  J^2 }{ N^2 \hbar^2 k_{B} T (e n_{s})^2} \intin d\om ~ \int_{0}^{2 k_{d}} q^2~ d q  Im \chi^{-+}(q,\om)(\frac{(- n(-\om))\om}{e^{\beta\hbar\om}-1})\nonumber\\ \label{h12a}
\eeqar
 To proceed further some assumption about $\chi^{-+}(q,\om)$ has to be made. We use Hamiltonian for $d$-eleectrons\cite{mathon,izuyama}
\beq
H_{d}=\sum_{\mathbf{k},\sigma} \ve(\mathbf{k})C^\dg_{\mathbf{k},\sigma}C_{\mathbf{k},\sigma}+\frac{\mathbb{I}}{\mathbb{N}}\sum_{q,\mathbf{k},\mathbf{k^\p},\sigma,\sigma^\p}C^\dg_{\mathbf{k}+q,\sigma}C_{\mathbf{k},\sigma}C^\dg_{\mathbf{k^\p},\sigma^\p}C_{\mathbf{k^\p}+q,\sigma^\p}
\eeq
 Here $\mathbb{I}$ is the exchange interaction parameter for d-band electrons, $\mathbb{N}$ is the number of lattice points. The susceptibility has been calculated in a famous paper by Izuyama et al.(1963)\cite{izuyama}. Here $C$ and $C^\dg$ are the annihilation and creation operators for d-electrons, $\ve(\mathbf{k})$ is  electron energy of Bloch state $k$ of d-electrons. Using random phase approximation the transverse susceptibility{IKK]\cite{izuyama} is given by
\beqar 
\chi^{-+}(q,\om)=\frac{\Gamma^{-+}(q,\om)}{1-\mathbb{I}\Gamma^{-+}(q,\om)}\label{chi1}
\eeqar 
where 
\beqar 
\Gamma^{-+}(q,\om)=\sum_{k} \frac{f(\ve_{k})-f(\ve_{k+q})}{\ve(k+q)-\ve(k)-\hbar\om} \label{gm1}
\eeqar
 $\Gamma^{-+}(q,\om)$ is the susceptibility of non interacting electrons, for which the band structure function $\ep_{k}$ is put to be constant. This can be seen if one's set $\mathbb{I}=0$. 
Employing identity $\lim_{\eta \to 0}\frac{1}{a\pm i \eta}=\mathfrak{p} (\frac{1}{a})\mp i\pi\delta (a)$ to equation (\ref{gm1}) the real part can be written as
\beq
\mathfrak{R}(q,\om)=\mathfrak{P}\sum_{k} \frac{f(\ve_{k})-f(\ve_{k+q})}{\ve(k+q)-\ve(k)-\hbar\om}  \label{chi2}
\eeq
Thus $\mathfrak{R}(q,\om)$ is the real part of non-interacting $d$-electrons susceptibility with wave number $q$ and frequency $\om$.
 $\mathfrak{I}(q,\om)$ is the corresponding imaginary susceptibility
\beq
\mathfrak{I}(q,\om)=\pi \sum_{k}f(\ve_{k})-f(\ve_{k+q}) \delta[\ve(k+q)-\ve(k)-\hbar\om] \label{chi3}
\eeq
Thus the electron spin susceptibility of d-electron in terms of real and imaginary part is written as
\beqar 
\Gamma^{-+}(q,\om)= \mathfrak{R}(q,\om)+i\mathfrak{I}(q,\om) \label{ap2a}
\eeqar
We note that $\mathfrak{R}(q,\om)$ is an even function of $q$. It can be expressed as
\beqar
\mathfrak{R}(q,\om)&=&\sum_{k}\frac{-q\frac{\pr f}{\pr k_{x}}-\frac{1}{2}(q.\nabla)^2 f|_{k}-\frac{1}{6}(q.\nabla)^3 f|_{k}}{q\frac{\pr \ep}{\pr k_{x}}+\frac{1}{2}(q.\nabla)^2\ve|_{k}-\hbar\om} \nonumber\\ \label{ap2}
\eeqar
The expression can be approximated to
\beq
\mathfrak{R}(q,\om)  \simeq  N(0)+\frac{N(0)}{4}(\frac{q\hbar}{2mv_{F}})^2 \simeq  N(0)+\frac{N(0)}{4} (\frac{\bar{q}}{2})^2\label{ap3}
\eeq
where $N(0)$ is the density of states at Fermi level, $E_{d}$  is energy of $d$-electrons. $\bar{q}=\frac{q}{k_{F}}$ is dimensionless wave-vector and $\hbar k_{F}=m v_{F}$.
\beq
\frac{\hbar^2 k_{F}q}{m}=v_{F}q\gg|\hbar\om|,  \label{cd1}
\eeq
and
\beq
\frac{q}{k_{F}}\ll1  \label{cd2}
\eeq
$k_{F}$ being the magnitude of wave number vector on the Fermi surface. The  The condition (\ref{cd2}) is applicabale since we are concerned with in the radius of fermi surface of s-electron and the condition (\ref{cd1}) comes from the fact that the energy change $\hbar\om$ of $s$-electron in a scattering process is order of $\frac{\hbar k_{F}q}{m_{s}}$ $(m_{s}$  is the mass of $s$-electrons). 
The imaginary part of susceptibility($\mathfrak{I}(q,\om))$ can be solved as
\beqar
\mathfrak{I}(q,\om)& = & \pi \sum_{k}(-q \frac{\partial f_{k}}{\partial k} \cos\phi)\delta\bigg(q\frac{\partial \ep_{k}}{\partial k}\cos\phi-\hbar\om\bigg) \nonumber\\
&=& \pi~ q \sum_{k}\bigg(-\frac{\partial f_{k}}{\partial \ve_{k}}\bigg)\frac{\partial \ve_{k}}{\partial k}\cos\phi \delta\bigg(q\frac{\partial \ve_{k}}{\partial k}\cos\phi-\hbar\om\bigg)  \label{h20}
\eeqar
Converting summation into integral as  $\sum_{k}=\frac{1}{2\pi}\int N(0)d\ve \int_{0}^{2\pi}d\phi$ and $N(0)$ defines the density of staes in two dimension . We replace $-\frac{\partial f_{k}}{\partial \ve_{k}}=\delta(\ve-\ve_{F})$ and first order drivatve of enegry with resprect to $k$ vector by $\hbar v_{F}$. The integral equation becomes as follows:
\beqar
\mathfrak{I} & = & \frac{q\hbar v_{F}}{2}\int_{0}^{\ve_{F}}N(0) \delta(\ve-\ve_{F}) d\ve\int_{0}^{2\pi}d\phi\cos\phi~  \delta\bigg(q\hbar v_{F}\cos\phi-\hbar\om \bigg) \nonumber\\ \label{h20a}
\eeqar
Using property of  delta function $ \int_{0}^{\ve_{F}}N(0)(\ve) \delta(\ve-\ve_{F})d\ve=\frac{N(0)}{2}$ for energy integral and using $\delta(ax)=\frac{1}{|a|}\delta(x)$ for $\phi$ integral, we get

\beqar
\mathfrak{I} & = &\frac{N(0)}{2} \int_{0}^{\pi}d\phi\cos\phi~  \delta\bigg(\cos\phi-\frac{\om}{q v_{F}} \bigg) \nonumber\\ \label{h20b}
\eeqar
Using  property $\delta(f(\theta))=\sum_{\theta_{0}}\frac{\delta(\theta-\theta_{0})}{|f^\p(\theta_{0})|}$, and setting $\theta=\cos^{-1}(\frac{\om}{qv_{F}})=\theta_{0}$, we obtain
\beqar
\mathfrak{I}=\frac{N(0)}{2}\frac{(\frac{\om}{q v_{F}})}{\sqrt{1-(\frac{\om}{q v_{F}})^2}}  \label{h22}
\eeqar
Apply the condition $\om<<q v_{F}$, the $\mathfrak{I}(q,\om)$ reduces to
\beqar
\mathfrak{I}=\frac{N(0)}{2}(\frac{\om}{q v_{F}}) = N(0)(\frac{\om k_{D}^{-1}}{2\bar{q} v_{F}})\label{h22a}
\eeqar
Here $\bar{q}=\frac{q}{k_{d}}$, $k_{d}$ is the d-electron Fermi vector, $\frac{\hbar^2 k^2_{d}}{2m_{d}}=E_{d}$ its fermi energy.

\beqar
\mathfrak{I}=\frac{N(0)}{4}\frac{\hbar~\om}{\bar{q}E_{d}}
\eeqar
Collecting the above information imaginary part of susceptibility (\ref{chi1})  reduce to

\beqar
Im\chi^{-+}(q,\om)=\frac{\mathfrak{I}}{(1-\mathbb{I}\mathfrak{R})^2+\mathbb{I}^2\mathfrak{I}^2} \label{im2}
\eeqar
Substituting real and imaginary part of transverse susceptibility from equations (\ref{ap7a}) and (\ref{h22}) into $Im\chi^{-+}(q,\om)$, we have
\beqar
Im\chi^{-+}(q,\om)=\frac{\frac{N(0)}{4}\frac{\hbar\om}{\bar{q}{E}_{d}}}{[1-(\mathbb{I}N(0)+\mathbb{I}N(0)(\frac{\bar{q}}{4})^2)]^2+\mathbb{I}^2[\frac{N(0)}{4}\frac{\hbar\om}{\bar{q}E_{d}}]^2} \label{h23}
\eeqar
Here $k_{0}^2=1-\mathbb{I}N(0)=1-\bar{\mathbb{I}}$ denotes the inverse of the RPA exchange enhnacement factor for $d$-band. We are interested in the behaviour of system for $k_{0}^2=0$ i.e. $c=c_{F}$. At $c=c_{F}$ one shift the critical point to a classical point to a desired low temperature regime. In other words nearness to a QCP is about by chemical doping\cite{mathon}. Here one can focus on the low temperature regime\cite{rice} for physical properties near the critical point. Therefore susceptibility takes the form:

\beqar
Im\chi^{-+}(q,\om)=\frac{\frac{N(0)}{4}\frac{\hbar\om}{\bar{q}{E}_{d}}}{[\frac{ \bar{\mathbb{I}}\bar{q}^2}{4^2}]^2+[\frac{\bar{\mathbb{I}}}{4}\frac{\hbar\om}{\bar{q}{E}_{d}}]^2}\label{h24}
\eeqar
\section{Result}
Writing $2d$ paramagnetic resistivity replacing $Im\chi^{-+}(q,\om)$ from equation (\ref{h24}) in expression (\ref{h12a}) we have
 \beqar
\rho^{para}_{2d}&=&\sqrt{\frac{m}{2\mu}}\frac{(12 \pi)  J^2 N(0) E_{d} }{ N^2\hbar^3 \bar{\mathbb{I}}^2  (e n_{s})^2 k_{d} k_{B} T}\int_{0}^{\frac{4{E}_{d}}{\hbar}} \frac{\om d\om}{(e^{\beta\hbar\om}-1)(1-e^{-\beta\hbar\om})}\int_{\frac{m_{d}\om}{\hbar k_{d}}}^{2k_{d}}  d q q^2\frac{\frac{q}{\om}}{[\frac{q^6}{k_{d}^6}(\frac{ {E}_{d}}{4\hbar\om})^2+1]} \nonumber\\&&
\eeqar

 put $t=\frac{q}{k_{d}}(\frac{{E}_{d}}{4\hbar\om})^{\frac{1}{3}}$, $q^3=t^3k_{d}^3(\frac{{E}_{d}}{4\hbar\om})^{-1}$ and write\\ 

prefactor $p_{0}=\sqrt{\frac{m}{2\mu}}\frac{(12\pi)  J^2 N(0) E_{d} }{ N^2 \hbar^3 \bar{\mathbb{I}}^2  (e n_{s})^2k_{d}}$, then
\beqar
\rho^{para}_{2d}&=&p_{0}k_{d}^4(\frac{4\hbar}{ E_{d}})^\frac{4}{3} \int_{0}^{\frac{4{E}_{d}}{\hbar}} \frac{\om^\frac{4}{3} d\om}{(e^{\beta\hbar\om}-1)(1-e^{-\beta\hbar\om})}\int_{l_{a}}^{l_{b}}\frac{t^3}{t^6+1},\nonumber\\&&
\eeqar
where limits for $t$-integral change to $l_{a}=\frac{1}{2}(\frac{\hbar\om}{2{E}_{d}})^\frac{2}{3}$ and $l_{b}=(\frac{2 {E}_{d}}{\hbar\om})^\frac{1}{3}$.\\
\vspace{3mm}
We put $\beta\hbar\om=\mathfrak{u}$ to make integrals temperature independent. Then resistivity simplifies to 

\beqar
\rho^{para}_{2d}&=& p_{0}k_{d}^4 (\frac{4 k_{B}T}{ {E}_{d}})^\frac{4}{3}\int_{0}^{\frac{4{E}_{d}}{k_{B}T}}\frac{ \mathfrak{u}^\frac{4}{3} d\mathfrak{u}}{(e^{\mathfrak{u}}-1)(1-e^{-\mathfrak{u}})}\int_{l^\p_{a}}^{l^\p_{b}}\frac{t^3}{t^6+1}\nonumber\\&&
\eeqar
where $l^\p_{a}=\frac{1}{2}(\frac{\mathfrak{u}k_{b}T}{2{E}_{d}})^\frac{2}{3}$, and $l^\p_{b}=(2 {E}_{d})^\frac{1}{3}(\mathfrak{u}k_{b}T)^\frac{-1}{3}$.\\
\vspace{3mm}
  And we have the final result $\rho^{para}_{2d}\propto T^\frac{4}{3}$

\section*{Conclusion}
We have performed a calculation for electrical resistivity in a $2d$ metal which is tuned near to its ferromagnetic instability. The required dynamical susceptibility in the expression of resistivity is calculated using Random Phase approximation. 
In 2$d$ we find that the real and imaginary parts of dynamical susceptibility are proportional to $q^2$ and $\frac{\om}{q}$ respectively.  This is similar to $3d$ case.  However, we find that the resistivity calculated for $2d$ case scales as $\rho_{para}^{2d} \propto T^\frac{4}{3}$, whereas in  $3d$ resistivity scales at $T^\frac{5}{3}$. Our result ($\rho_{2d}^{para} \propto T^\frac{4}{3}$) agrees with the SCR theory.
\appendix
\section{Appendix: mathematical details of $\phi$ intergral}
To solve the term  $\int_{0}^{2\pi} d\phi ~\delta(\cos\phi+\underbrace{ \frac{m\om}{\hbar k q}+\frac{q}{2 k}}_{f(k,q,\om)})$.  We use delta function property $ \delta(F(x))=\sum_{i}\frac{\delta(x-x_{i})}{|F^\p(x_{i})|}$. We can put $\cos\phi+ \frac{m\om}{\hbar k q}+\frac{q}{2 k}=0$, $\phi=\cos^{-1}( -\frac{m\om}{\hbar k q}-\frac{q}{2 k})=\phi_{0}(q,k,\om)$ and here $\frac{m\om}{\hbar  q}+\frac{q}{2 }<k$.
\beqar
\int_{0}^{2\pi} d\phi ~\delta(\cos\phi+ f(k,q,\om))&=&\int_{0}^{2\pi} d\phi \frac{\delta(\phi-\phi_{0}(q,k,\om))}{|-\sin\phi|_{\phi_{0}}}\nonumber\\
&=& \frac{1}{\sin\phi_{0}}  \int_{0}^{2\pi} d\phi \delta(\phi-\phi_{0})
=\frac{1}{\sin\phi_{0}}\nonumber\\
&=& \frac{1}{\sqrt{1-(\frac{m\om}{\hbar k  q}+\frac{q}{2 k })^2}}=\frac{1}{\sqrt{k^2-\underbrace{(\frac{m\om}{\hbar q}+\frac{q}{2  })^2}_{q_{0}^2}}}
\label{h14} 
\eeqar

\section{ mathematical details of $k$ intergral}

\beqar
I_{k}&=&\int_{q_{0}}^{\infty} k~ d k  \frac{f^{0}(\epsilon_{k})(1-f^{0}(\ep_{k}-\hbar \om))}{\sqrt{k^2-q_{0}^2}}  \label{h13b}
\eeqar
Converting k integral into energy  $k=\frac{\sqrt{2m\ep}}{\hbar}$, $dk=\frac{\sqrt{m}d\ep}{\hbar\sqrt{2\ep}}$ and writing lower limit for energy  $\ep_{0}=\frac{\hbar^2 q_{0}^2}{2 m }$ and upperlimit for enegy integral becomes infinite.
\beqar
I_{\ep}&=& \frac{m}{\hbar^2}\int_{\ep_{0}}^{\infty} d\ep \frac{f^{0}(\epsilon_{k})(1-f^{0}(\ep_{k}-\hbar \om))}{\sqrt{\frac{2m\ep}{\hbar^2}-q_{0}^2}}\nonumber\\
\eeqar
Replacing the value Fermi function $f^{0}(\epsilon_{k})=\frac{1}{e^{\beta(\ep-\mu)}+1}$ 
\beqar
I_{\ep}&=& \frac{\sqrt{m}}{\hbar\sqrt{2}}\int_{\ep_{0}}^{\infty} \frac{d\ep}{\sqrt{\ep-\frac{\hbar^2 q_{0}^2}{2 m}}}(\frac{1}{e^{\beta(\ep-\mu)}+1})\bigg(1-\frac{1}{e^{\beta(\ep-\mu-\hbar\om)}+1}\bigg)\nonumber\\
&=& \frac{1}{\hbar}\sqrt{\frac{m}{2}}\int_{\ep_{0}}^{\infty} \frac{d\ep}{\sqrt{\ep-\ep_{0}}} (\frac{1}{e^{\beta(\ep-\mu)}+1})
\bigg(\frac{e^{\beta(\ep-\mu-\hbar\om)}}{e^{\beta(\ep-\mu-\hbar\om)}+1}\bigg)    \label{h14a}
\eeqar
 To simplify it further we take $\alpha=e^{-\beta\hbar\om}$, $u=e^{\beta(\ep-\mu)}$, and its $\log$ gives $\ep=\mu+\frac{1}{\beta}\log u$. In low temperature case if we apply condition $\ep_{0}\ll\mu$ the lower limit for $u$ becomes zero and higher limit goes to infinity. The above expression converts in new form as

\beqar
I_{\ep}&=&  \frac{\alpha}{\beta\hbar}\sqrt{\frac{m}{2}}\int_{0}^{\infty}\frac{du}{\sqrt{\mu+\frac{1}{\beta}\log u-\ep_{0}}}(\frac{1}{u+1})(\frac{1}{\alpha u+1})  \label{h14b}
\eeqar
Using the above defined  condition of low temperature limit, we can reduce the square root term as $\sqrt{\mu+\frac{1}{\beta}\log u-\ep_{0}}\simeq \sqrt{\mu}$.

\beqar
I_{\ep}&=&  \frac{\alpha}{\beta\hbar}\sqrt{\frac{m}{2\mu}}\int_{0}^{\infty} \frac{du}{(u+1)(\alpha u+1)}   \label{h14c}
\eeqar
This elementary integral reduces to
\beqar
I_{\ep}&=&  \frac{\alpha}{\beta\hbar}\sqrt{\frac{m}{2\mu}}\bigg[\int_{0}^{\infty}\frac{du}{(u+1)(1-\alpha)}+\frac{\alpha}{\alpha-1}\int_{0}^{\infty}\frac{du}{\alpha u+1}\bigg] \nonumber\\
&=& \frac{\alpha}{\beta\hbar}\sqrt{\frac{m}{2\mu}}(\frac{1}{1-\alpha})\log|\frac{1}{\alpha}|  \label{h15}
\eeqar

writing $\alpha=e^{-\beta\hbar\om}$, we have
\beqar
I_{\ep}&=& \frac{\alpha}{\beta\hbar}\sqrt{\frac{m}{2\mu}}\bigg(\frac{-\log (e^{-\beta\hbar\om})}{1-e^{-\beta\hbar\om}}\bigg) = \sqrt{\frac{m}{2\mu}}(\frac{\om}{e^{\beta\hbar\om}-1}) \label{h15a}
\eeqar

\section{ mathematical details of real part of susceptibility}

\beqar
\mathfrak{R}(q,\om)&=&\sum_{k}\frac{-(q.\nabla) f|_{k}-\frac{1}{2}(q.\nabla)^2 f|_{k}-\frac{1}{6}(q.\nabla)^3 f|_{k}}{q\frac{\pr \ep}{\pr k_{x}}+\frac{1}{2}(q.\nabla)^2\ve|_{k}-\hbar\om} \nonumber\\ \label{ap2}
\eeqar
simplifies to
\beqar
\mathfrak{R}(q,\om)=\sum_{k} \frac{-q\frac{\pr f}{\pr \ep}\frac{\pr \ep}{\pr k}\frac{\pr k}{\pr k_{x}}-\frac{q^2}{2}\frac{\pr^2 f}{\pr k_{x}^2}-\frac{q^2}{6}\frac{\pr^3 f}{\pr k_{x}^2}}{q\frac{\pr \ep}{\pr k_{x}}+\frac{q^2}{2}\frac{\pr^2 \ep}{\pr k_{x}^2}-\hbar\om} \label{real1}
\eeqar

writing $\frac{\pr f}{\pr k_{x}}=\frac{\pr f}{\pr \ep}\frac{\pr \ep}{\pr k}\frac{\pr k}{\pr k_{x}}=-\hbar v_{F}\delta(\ep-\ep_{F})\cos\theta$ and $\frac{k_{x}}{k}=\cos\theta$.  The higher derivtives of fermi function can be written in the form $\frac{\pr^2 f}{\pr k_{x}^2}=\frac{\pr}{\pr k_{x}}(\frac{\pr f}{\pr k_{x}})=\frac{\pr}{\pr k_{x}}(\frac{\pr f}{\pr \ep}\frac{\pr \ep}{\pr k}\frac{\pr k}{\pr k_{x}})$, therefore the double derivtaive of fermi function becomes $-(\hbar v_{F})^2 \cos^2\theta\delta^\p(\ep-\ep_{F})$. The real part takes the form 

\beqar
\mathfrak{R}(q,\om)=\sum_{k}\frac{q\hbar v_{F}\delta(\ep-\ep_{F})\cos\theta+\frac{q^2}{2}(\hbar v_{F})^2\delta^\p(\ep-\ep_{F})+\frac{q^3}{6}(\hbar v_{F})^3\delta^{\p\p}(\ep-\ep_{F})}{q\hbar v_{F}\cos\theta+\frac{\hbar^2 q^2}{2m}\cos^2\theta-\hbar\om} \label{real2}
\eeqar
Converting sum into integral, we have
\beqar
\mathfrak{R}(q,\om)&=&\frac{1}{\pi}\bigg[\int_{0}^{\infty} N_{2d}(0)\delta(\ep-\ep_{F})d\ep \int_{0}^{2\pi}\frac{ \cos\theta d\theta}{\cos\theta+\frac{q\hbar}{2mv_{F}}\cos^2\theta-\frac{\hbar\om}{\hbar q v_{F}}} +\frac{q\hbar v_{F}}{2}\times\nonumber\\ &&~~~~~~\int_{0}^{\infty}  N_{2d}(0)\delta^\p(\ep-\ep_{F})d\ep\int_{0}^{2\pi}\frac{ \cos^2\theta d\theta}{\cos\theta+\frac{q\hbar}{2mv_{F}}\cos^2\theta-\frac{\hbar\om}{\hbar q v_{F}}} +\frac{(q\hbar v_{F})^2}{6} \times\nonumber\\&&~~~~~~~\int_{0}^{\infty}  N_{2d}(0)\delta^{\p\p}(\ep-\ep_{F})d\ep \int_{0}^{2\pi}\frac{ \cos^3\theta d\theta}{\cos\theta+\frac{q\hbar}{2mv_{F}}\cos^2\theta-\frac{\hbar\om}{\hbar q v_{F}}}\bigg] \label{real3}
\eeqar
the expression further gives
\beqar
\mathfrak{R}(q,\om)=\frac{1}{2\pi}N(0) \int_{0}^{2\pi}\frac{ \cos\theta d\theta}{\cos\theta+\frac{q\hbar}{2mv_{F}}\cos^2\theta-\frac{\hbar\om}{\hbar q v_{F}}} \label{real4}
\eeqar
put $\theta=\phi+\pi$, $\alpha=\frac{q\hbar}{2mv_{F}}$ and $\beta=\frac{\hbar\om}{\hbar q v_{F}}$, the above integral becomes
\beqar
\mathfrak{R}(q,\om)=\frac{1}{\pi}N(0) \underbrace{\int_{0}^{\pi}\frac{ \cos\phi d\phi}{\cos\phi-\alpha\cos^2\phi+\beta}}_{I^\p} \label{real5}
\eeqar
Set $\cos\phi=x$, thus we obtain

\beqar
\mathfrak{R}(q,\om)=\frac{N(0) }{\pi} (\frac{1}{-\alpha})\int_{-1}^{1}\frac{d x}{\sqrt{1-x^2}}\frac{x}{[(x-\frac{1}{2\alpha})^2-\gamma^2]} \label{real6}
\eeqar
 here $\gamma=(\frac{\beta}{\alpha}+\frac{1}{4\alpha^2})^\frac{1}{2}$.  The integral reduces to
\beqar
\mathfrak{R}(q,\om)=\frac{N(0) }{\pi} (\frac{1}{-\alpha})\int_{-1}^{1}\frac{d x}{\sqrt{1-x^2}}\bigg[\frac{\eta}{\eta+\eta^\p}(\frac{1}{x-\eta})+\frac{\eta^\p}{\eta+\eta^\p}(\frac{1}{x+\eta^\p})\bigg] \label{real7}
\eeqar

where $\eta=\frac{1}{2\alpha}+(\frac{\beta}{\alpha}+\frac{1}{4\alpha^2})^\frac{1}{2}$ and $\eta^\p=-\frac{1}{2\alpha}+(\frac{\beta}{\alpha}+\frac{1}{4\alpha^2})^\frac{1}{2}$ are q dependent parameters. In limit $q\rta0$, $\alpha$ and $\beta$  are very small values, so that $\eta>>1$  and $\eta^\p<<1$. Thus we have

\beqar
\mathfrak{R}(q,\om)&=&\frac{N(0) }{\pi} (\frac{1}{-\alpha})\frac{\eta}{\eta+\eta^\p}\int_{-1}^{1}\frac{d x}{\sqrt{1-x^2}}(-\frac{1}{\eta})\frac{1}{(1-\frac{x}{\eta})}\nonumber\\&& \label{real8}
\eeqar
Using series expansion method to solve integral, we get
\beqar
\mathfrak{R}(q,\om)&=& \frac{N(0) }{\pi}\frac{(\eta+\eta^\p)^{-1}}{\alpha}\bigg[\int_{-1}^{1}\frac{d x}{\sqrt{1-x^2}}+\frac{1}{\eta}\int_{-1}^{1}\frac{xd x}{\sqrt{1-x^2}}+\frac{1}{2\eta^2}\int_{-1}^{1}\frac{x^2d x}{\sqrt{1-x^2}}\bigg]\nonumber\\&&  \label{real9}
\eeqar
for small $\alpha$ and $\beta$ sum of $\eta$ and $\eta^\p$ gives $(\alpha )^{-1}$. To solve integral terms we set $\cos\theta=x$, which reduces the expression as
\beqar
\mathfrak{R}(q,\om)&=& \frac{N(0) }{\pi}[\pi+\frac{\pi}{4\eta^2}] \label{real10}
\eeqar
setting $\eta=\frac{1}{\alpha}=\frac{2mv_{F}}{q\hbar}$, the real part of susceptibility becomes
\beqar
\mathfrak{R}(q,\om)=N(0) +\frac{N(0) }{4}(\frac{q\hbar}{2mv_{F}})^2 \label{real11}
\eeqar

\section*{Acknowledgement}
I (Komal Kumari) thank  Physical Research Laboratory(PRL), for providing me local hospitality during this work. We thank Prof. Dietrich Belitz for clarifying the importance of  first order transitions in clean magnetic metals, and importance of disorder in second order transitions.


\end{document}